\documentclass[sigconf]{acmart}

\AtBeginDocument{%
  }

\copyrightyear{2023} 
\acmYear{2023} 
\setcopyright{acmlicensed}\acmConference[WWW '23]{Proceedings of the ACM Web Conference 2023}{May 1--5, 2023}{Austin, TX, USA}
\acmBooktitle{Proceedings of the ACM Web Conference 2023 (WWW '23), May 1--5, 2023, Austin, TX, USA}
\acmPrice{15.00}
\acmDOI{10.1145/3543507.3583421}
\acmISBN{978-1-4503-9416-1/23/04}

\usepackage[utf8]{inputenc}

\usepackage{CJKutf8}
\usepackage{microtype}
\usepackage{algorithm}
\usepackage{booktabs} 
\usepackage{graphicx}
\usepackage{amsfonts}
\usepackage{amsmath}
\usepackage{subfigure}
\usepackage[T1]{fontenc}
\usepackage{makecell}
\usepackage{multirow}
\usepackage{diagbox}
\usepackage{amsthm,amsmath}

\usepackage{amssymb}
\usepackage{multicol,multirow}
\usepackage{mathrsfs}

\usepackage{amsmath,bm}
\usepackage{amsfonts}

\def\eqref#1{equation~\ref{#1}}

\def\1{\bm{1}}

\def\vd{{\bm{d}}}

\def\vp{{\bm{p}}}
\def\vq{{\bm{q}}}

\def\vw{{\bm{w}}}

\DeclareMathAlphabet{\mathsfit}{\encodingdefault}{\sfdefault}{m}{sl}
\SetMathAlphabet{\mathsfit}{bold}{\encodingdefault}{\sfdefault}{bx}{n}

\def\sD{{\mathbb{D}}}

\def\sP{{\mathbb{P}}}
\def\sQ{{\mathbb{Q}}}

\let\Ls\relax
\newcommand{\Ls}{\mathcal{L}}
\let\KL\relax
\newcommand{\KL}{D_{\mathrm{KL}}}

\usepackage[noabbrev,capitalize]{cleveref}

\crefname{equation}{equation}{equations}
\crefname{footnote}{footnote}{footnotes}
\crefname{line}{line}{lines}
\Crefname{section}{\S}{\S\S}
\crefformat{section}{#2\S#1#3}
\crefrangeformat{section}{\S\S#3#1#4--#5#2#6}

\begin{document}
\fancyhead{}

\title{PROD: Progressive Distillation for Dense Retrieval}

\author{
Zhenghao Lin$^{1}$\orcid{0000-0001-9172-1628},
Yeyun Gong$^{2,*}$\orcid{0000-0001-9954-9674},
Xiao Liu$^{2,*}$\orcid{0000-0002-8893-366X},
Hang Zhang$^{2,*}$\orcid{0000-0002-9940-3517},
Chen Lin$^{1,\dagger}$\orcid{0000-0002-2275-997X},\\
Anlei Dong$^{3}$\orcid{0000-0002-8241-4746},
Jian Jiao$^{3}$\orcid{0000-0003-4779-9588},
Jingwen Lu$^{3}$\orcid{0000-0001-8208-898X},
Daxin Jiang$^{3}$\orcid{0000-0002-6657-5806},
Rangan Majumder$^{3}$\orcid{0000-0003-2430-575X},
Nan Duan$^{2}$\orcid{0000-0002-3387-4674}
}
\def \authors{Zhenghao Lin, Yeyun Gong, Xiao Liu, Hang Zhang, Chen Lin, Anlei Dong, Jian Jiao, Jingwen Lu, Daxin Jiang, Rangan Majumder, Nan Duan}
\affiliation{%
  \institution{$^1$School of Informatics, Xiamen University, $^2$Microsoft Research Asia, $^3$Microsoft}
  \city{}
  \country{}
}
\email{zhenghaolin@stu.xmu.edu.cn, chenlin@xmu.edu.cn}
\email{{yegong,xiaoliu2,v-zhhang,nanduan}@microsoft.com}
\thanks{$^{\dagger}$Chen Lin is the corresponding author.}
\thanks{$^{*}$Equal contribution.}

\renewcommand{\shortauthors}{Lin et al.}

\begin{abstract}
Knowledge distillation is an effective way to transfer knowledge from a strong teacher to an efficient student model.
Ideally, we expect the better the teacher is, the better the student performs.
However, this expectation does not always come true.
It is common that a strong teacher model results in a bad student via distillation due to the nonnegligible gap between teacher and student.
To bridge the gap, we propose~\textbf{PROD}, a~\textbf{PRO}gressive~\textbf{D}istillation method, for dense retrieval.
PROD consists of a \textit{teacher progressive distillation} and a \textit{data progressive distillation} to gradually improve the student. To alleviate catastrophic forgetting, we introduce a regularization term in each distillation process.
We conduct extensive experiments on seven datasets including five widely-used publicly available benchmarks: MS MARCO Passage, TREC Passage 19, TREC Document 19, MS MARCO Document, and Natural Questions, as well as two industry datasets: \textsc{Bing-Rel} and \textsc{Bing-Ads}. PROD achieves the state-of-the-art in the distillation methods for dense retrieval. Our 6-layer student model even surpasses most of the existing 12-layer models on all five public benchmarks. 
The code and models are released in \href{https://github.com/microsoft/SimXNS}{https://github.com/microsoft/SimXNS}.
\end{abstract}



\maketitle

\section{Introduction}
In recent years, pre-trained models have made breakthroughs in various NLP tasks, including question answering~\cite{devlin2018bert}, summarization~\cite{lewis2019bart,qi2020prophetnet}, and dense retrieval~\cite{DBLP:conf/emnlp/KarpukhinOMLWEC20,DBLP:conf/iclr/ZhangGS0DC22}.
To further improve the performance of end tasks, large models are proposed~\cite{brown2020language,smith2022using}.
Despite their successful applications on small or medium scale benchmarks, the efficiency issue of model inference becomes a problem. 
In the practical scenarios of dense retrieval, online systems need to retrieve the relevant documents from a large number of candidates, and answer the user queries in time. 
Therefore, an efficient small model is particularly critical in dense retrieval applications.

To take into account both performance and efficiency, knowledge distillation techniques have been widely used~\cite{sanh2019distilbert,jiao2019tinybert,DBLP:journals/corr/abs-2205-09153}.
In previous work, \citet{DBLP:conf/sigir/ZengZV22} proposed a curriculum learning method for dense retrieval distillation.
\citet{DBLP:conf/emnlp/RenQLZSWWW21} and \citet{DBLP:conf/iclr/ZhangGS0DC22} proposed the re-ranker as teacher method.
\citet{lin2021batch} proposed an in-batch negative distillation method with ColBERT~\cite{DBLP:conf/sigir/KhattabZ20}.
\citet{DBLP:journals/corr/abs-2205-09153} proposed interaction distillation, cascade distillation, and dual regularization to bring the re-ranker with retriever.
These methods demonstrate the importance of knowledge distillation in dense retrieval tasks from different perspectives.
However, when the gap between student and teacher is very large, how to better close the gap is a big challenge.
Both the experiments in the previous NLG task~\cite{DBLP:conf/iclr/ZhouGN20} and our experiments in the dense retrieval task provide ample support for the proposition that \emph{the performance of the teacher and the student is not positively correlated}.

To solve this problem, there is a branch of work utilizing progressive distillation, in which teachers are dynamically adjusted to distill students.
Apart from refining the pretraining workflow \cite{DBLP:journals/corr/abs-2106-02241}, progressive distillation methods have achieved great success in various down-stream tasks in NLP and CV \cite{DBLP:journals/corr/abs-2110-08532,DBLP:conf/iclr/SalimansH22,DBLP:conf/acl/HuangXYWCLCXRLD22,DBLP:conf/cvpr/AndonianCH22}.
The main idea of the existing progressive distillation work is to use stronger teachers while educating students.
However, there are two shortcomings.
First, this idea has not been proven to be useful in dense retrieval yet.
Second, the existing studies \cite{DBLP:journals/corr/abs-2106-02241,DBLP:journals/corr/abs-2110-08532,DBLP:conf/iclr/SalimansH22,DBLP:conf/acl/HuangXYWCLCXRLD22,DBLP:conf/cvpr/AndonianCH22} mainly focus on teacher models, neglecting the importance of training data in the process of progressive distillation.

In this paper, we propose a progressive distillation method, PROD, to minimize the gap between the teacher and the student.
PROD consists of two progressive mechanisms: \textit{teacher progressive distillation} (TPD) and \textit{data progressive distillation} (DPD).
In TPD, we gradually improve the capability of teachers by using different architectures, enabling student model to learn knowledge progressively.
In DPD, we start to let students learn from all the data, and then gradually select samples that the student is confused about for strengthening.
In each progressive step, we introduce regularization loss to avoid catastrophic forgetting of the knowledge memorized in the previous step.

The motivation of PROD is from two aspects:
1) When the gap between teacher and student is huge, a stronger teacher is not necessarily better than a relatively weak teacher.
Such as, the university professor may not be more suitable than a kindergarten teacher to teach a kindergarten student.
Therefore, we design TPD to enhance teachers gradually.
2) There are different knowledge suitable for the student model to learn at different stages, such as middle school textbooks are suitable for middle school students to learn.
Thus, we design DPD to select the appropriate (not too easy or hard) knowledge for the student to learn.

We conduct extensive experiments on five widely-used benchmarks (MS MARCO Passage, TREC Passage 19, TREC Document 19, MS MARCO Document, and Natural Questions) and two industry datasets (\textsc{Bing-Rel} and \textsc{Bing-Ads}).
The results of extensive experiments on five popular benchmarks show the effectiveness of PROD, and performance on two industry datasets also demonstrates the commercial value of PROD.

\section{Related work}
\label{sec:related_work}

This work is related to two lines of work.
\subsection{Dense Retrieval}
Compared with sparse retrieval methods \cite{DBLP:conf/sigir/Yang0L17,DBLP:conf/sigir/DaiC19,DBLP:journals/corr/abs-1904-08375}, dense retrieval has the potential to find hidden semantic correlations between queries and passages.
Several directions have been explored to improve the performances of the popular dual encoder structure, including finding hard negatives with higher qualities \cite{DBLP:conf/emnlp/KarpukhinOMLWEC20,DBLP:conf/iclr/XiongXLTLBAO21,DBLP:conf/naacl/QuDLLRZDWW21}, multi-vector interactions \cite{DBLP:conf/sigir/KhattabZ20}, and the joint training of retrievers and re-rankers \cite{DBLP:conf/emnlp/RenQLZSWWW21,DBLP:conf/iclr/ZhangGS0DC22}.

Meanwhile, another line of the work studied knowledge distillation for dense retrieval by using a single teacher model \cite{DBLP:conf/emnlp/RenQLZSWWW21, DBLP:journals/corr/abs-2208-13661}, multiple teacher models with joint training techniques \cite{DBLP:journals/corr/abs-2205-09153} and curriculum learning \cite{DBLP:conf/sigir/ZengZV22}.
Our work is in line with these work, while having the key difference that we choose one teacher model with progressive ability level and focusing on the confusing data with progressive difficulty in each training stage.
\subsection{Knowledge Distillation}
The knowledge distillation \cite{DBLP:journals/corr/HintonVD15} has been widely studied for decades.
There are a variety of off-the-shelf practical techniques, such as the response-based method \cite{DBLP:conf/ijcai/KimOKCY21}, the feature-based method \cite{DBLP:journals/corr/RomeroBKCGB14} and distilling the attention scores \cite{DBLP:conf/iclr/ZagoruykoK17} with Kullback–Leibler divergence (KLD) or mean squared error (MSE).
Recently, a rising group of work focuses on the relationships between the teacher model and the student model.
Some elaborate the knowledge transmission methods \cite{DBLP:conf/cvpr/Chen0ZJ21,DBLP:conf/acl/ZhouXM22}, some introduce middle models \cite{DBLP:conf/aaai/MirzadehFLLMG20}, and others work on distillation with multiple teachers \cite{DBLP:journals/corr/abs-2204-00548}.
It can be noted that how to use a more reasonable teacher model to guide the learning of the student model has gradually become the current key direction.

Among the studies on knowledge distillation, there is a branch of work on progressive distillation, in which teachers are dynamically adjusted to distill students.
In spite of the success on the workflow of pretraining \cite{DBLP:journals/corr/abs-2106-02241}, progressive distillation methods are widely proven to be effective in down-stream tasks, such as image classification \cite{DBLP:journals/corr/abs-2110-08532}, image generation \cite{DBLP:conf/iclr/SalimansH22}, GLUE \cite{DBLP:journals/corr/abs-2110-08532,DBLP:conf/acl/HuangXYWCLCXRLD22}, question answering \cite{DBLP:journals/corr/abs-2110-08532}, and cross-modal representation learning \cite{DBLP:conf/cvpr/AndonianCH22}.
In each task, the main idea of the existing progressive distillation work is to use stronger teachers while educating students. 
Different from the above studies, PROD is adopt to the research field of dense retrieval, using a progressive method from two perspectives, teacher and data, in which teachers with different architectures are applied in a progressive order and more and more confusing data is mined to fill the performance gap between the teacher and the student.

\section{Preliminary}

\begin{figure*}[t]
\centering
\includegraphics[width=0.73\textwidth]{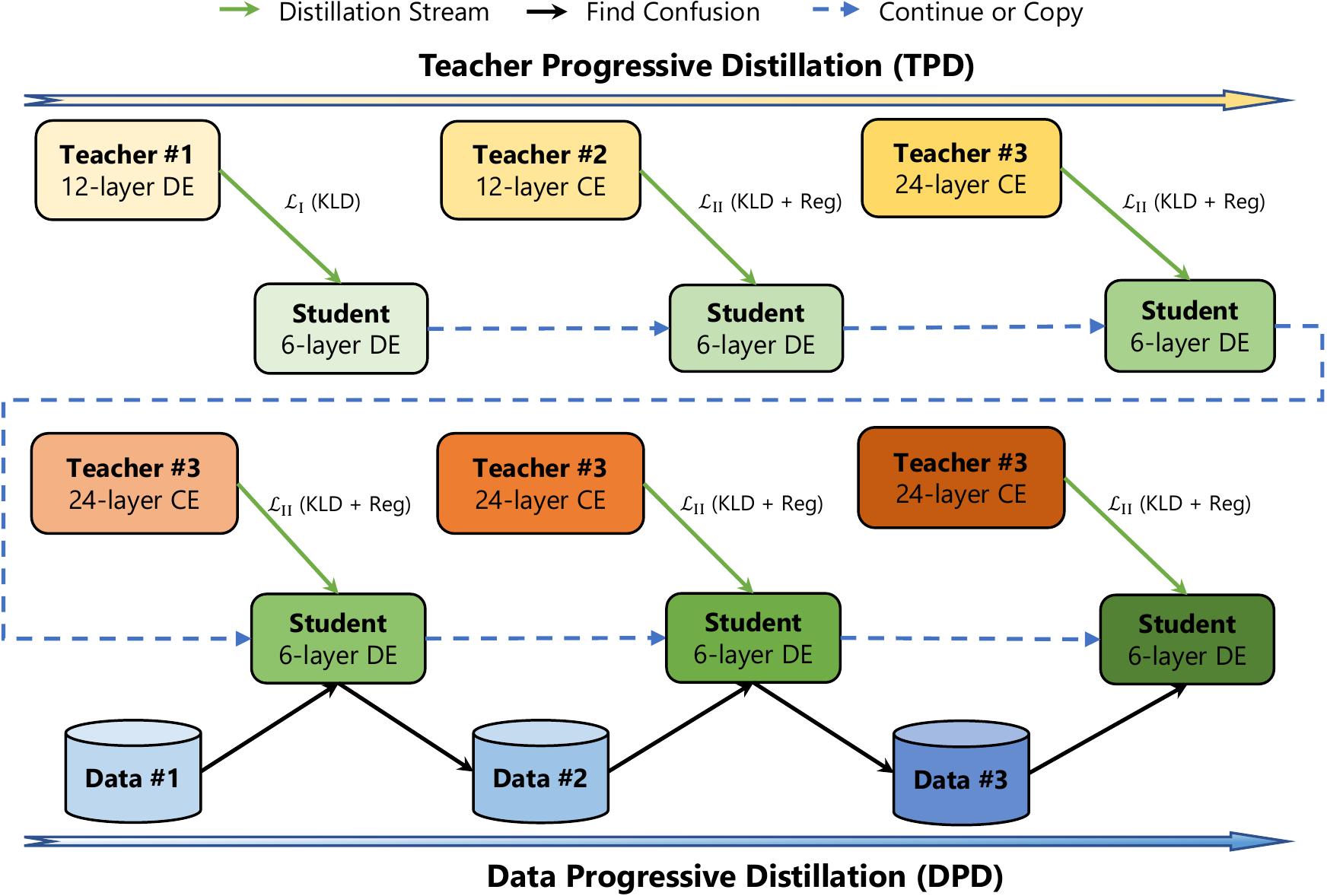}
\caption{The framework of PROD. In the TPD, we used three types of teachers, namely, 12-layer DE, 12-layer CE, and 24-layer CE. In the DPD, we used 24-layer CE to mine confusing data and iterative distillation training. Each step of distillation is continued with the 6-layer DE student in the previous step. Furthermore, to alleviate catastrophic forgetting, we employ a regularization loss item to maintain the stability of distillation.}
\label{fig:pipeline}
\end{figure*}

\subsection{Task Description}

We follow the basic definition of dense text retrieval \cite{DBLP:conf/emnlp/KarpukhinOMLWEC20,DBLP:conf/iclr/XiongXLTLBAO21,DBLP:conf/iclr/ZhangGS0DC22,DBLP:conf/sigir/KhattabZ20}.
Assume we have a query set $\sQ=\{ \vq_1, \vq_2, \dots, \vq_{n} \}$ containing $n$ queries and a passage set $\sP=\{ \vp_1, \vp_2, \dots, \vp_{m} \}$, our target is to find the most relevant passages in corpus $\sP$ for each query $\vq_i$.

\subsection{Dual Encoder}

The currently popular architecture for dense retrieval is dual encoder (DE), which can encode queries and passages into dense vectors $\vp$ and $\vq$, respectively, and calculate the similarity scores through the inner product as:
\begin{equation}
\footnotesize
s_{de}(\vq, \vp)=E_{Q}(\vq)^{T} \cdot E_{P}(\vp)
\label{equ:descore}
\end{equation}
where $E_{Q}(\cdot)$ and $E_{P}(\cdot)$ are the query encoder and the passage encoder, respectively.
Based on the embedding vectors, existing solutions generally employ approximate nearest neighbor (ANN) search algorithms like faiss \cite{DBLP:journals/tbd/JohnsonDJ21}.

\subsection{Cross Encoder}

In order to capture the fine-grained relationships between a pair of query and passage, the cross encoder (CE) is often used as the re-ranking model, rearranging the retriever outputs to improve the retrieval performances.
In particular, the concatenation of a query $\vq$ and a passage $\vp$ are with the special token \texttt{[SEP]} is the input of CE.
The similarity score is calculated by the \texttt{[CLS]} embedding vector of CE as:
\begin{equation}
\footnotesize
s_{ce}(\vq, \vp)=\vw^T \cdot E_{ce}([\vq;\vp])
\label{equ:cescore}
\end{equation}
where $E_{ce}(\cdot)$ is an encoder that can be initialized with any pre-trained language models, $[;]$ is concatenation operation, and $\vw$ is the linear projector.

\subsection{Knowledge Distillation}

Knowledge distillation is a simple yet effective approach to boost a small model (student model) with the additional knowledge from a big model (teacher model) \cite{DBLP:journals/corr/HintonVD15}.
In the task of dense retrieval, the student model is usually trained with \textbf{hard labels} such as the relevance annotations for each query-passage pair using a contrastive loss.
Additionally, it can also benefit from comparing with the prediction distribution of the teacher model, \textit{aka} \textbf{soft labels}, by decreasing a measurement like KLD and MSE, indicating the differences between the prediction distributions \cite{DBLP:conf/emnlp/RenQLZSWWW21,DBLP:conf/naacl/SanthanamKSPZ22,DBLP:journals/corr/abs-2205-09153}.

\section{Method}

We show the framework of the progressive distillation method, namely PROD, in \cref{fig:pipeline}.
The main idea of PROD is to gradually minimize the gap between a well-trained teacher model (24-layer CE) and the target student model (6-layer DE) by two sequential procedures, which are called \textit{teacher progressive distillation} (TPD) and \textit{data progressive distillation} (DPD).

\subsection{Teacher Progressive Distillation}
\label{sec:teacher_progressive}

Previous work found that different data instances would have various difficulties for a particular family of models to learn \cite{DBLP:conf/emnlp/SwayamdiptaSLWH20,DBLP:conf/icml/EthayarajhCS22}.
Similarly, we believe that the probability distributions of different teacher models when selecting the relevant passages from a candidate pool are also different.
Therefore, when facing the various difficulties of data instances in training, we use three different teacher models with different ability levels to civilize the student model gradually.
More specifically, we employ three progressive teacher models, \textit{i.e.}, a 12-layer DE, a 12-layer CE and a 24-layer CE, to boost a 6-layer DE student model\footnote{We use the notation ``X $\rightarrow$ Y'' to denote the distillation stage of a teacher model X and a student model Y.} in order.
Before learning from a teacher model, we retrieve the top-$k$ negatives \cite{DBLP:conf/iclr/XiongXLTLBAO21} with the current student checkpoint, randomly sample a subset and train a teacher model using the mined hard negatives.

\paragraph{Distilling with a DE Teacher.}
We first use the original data to warm up the 12-layer teacher DE and the 6-layer student DE, and then use the isomorphic distillation, \textit{i.e.}, 12-layer DE $\rightarrow$ 6-layer DE.
The loss function can be divided into two parts.

The first part is calculated by the output distribution of the student model and hard labels, which is called \textbf{hard loss}:
\begin{equation}
\footnotesize
\Ls_{h}(\vq,\vp^+,\sP^-) = -\log\frac{\exp(s_{de}^S(\vq, \vp^+))}{\sum_{\vp \in \{\vp^+, \sP^-\}} \exp(s_{de}^S(\vq, \vp))}
\label{equ:dtdhardloss}
\end{equation}
where $\vp^+$ and $\sP^-$ is the relevant passage and negative passage pool of $\vq$, respectively;
$s_{de}^{S}$ is the similarity scores of the student model.

The other part measures the differences of the probability distributions between the teacher model and the student model given the same batch of data, which is called \textbf{soft loss}:
\begin{equation}
\footnotesize
\vd^S_{de} = \frac{\exp(s_{de}^{S}(\vq, \vp))}{\sum_{\vp' \in \{\vp^+, \sP^-\}}\exp(s_{de}^{S}(\vq, \vp'))} \label{equ:studentscorepd}
\end{equation}
\begin{equation}
\footnotesize
\vd^T_{de} = \frac{\exp(s_{de}^{T}(\vq, \vp))}{\sum_{\vp' \in \{\vp^+, \sP^-\}}\exp(s_{de}^{T}(\vq, \vp'))} \label{equ:deteacherscorepd}
\end{equation}
\begin{equation}
\footnotesize
\Ls_{de\rightarrow de}(\vq,\vp^+,\sP^-)=\KL(\vd^T_{de}/\tau, \vd^S_{de}/\tau) \label{equ:dtdsoftloss}
\end{equation}
where $s_{de}^{T}$ is the similarity scores of the teacher model;
$\tau$ is the temperature of distillation.
At last, the final distillation loss with a DE teacher is a weighted sum of the before-mentioned two parts: 
\begin{equation}
\footnotesize
\Ls_{\textrm{I}} = \alpha_1 \Ls_{h} + \beta_1 \Ls_{de \rightarrow de}
\label{equ:weightedloss}
\end{equation}
where $\alpha_1$ and $\beta_1$ control the importance of hard loss and soft loss.

There are two important reasons why we use isomorphic distillation.
(1) First, according to previous research on DE \cite{DBLP:conf/emnlp/KarpukhinOMLWEC20}, adding more hard negatives when training DE can not lead to further improvements.
Therefore, simply training a DE may be not suitable for difficult data instances.
Meanwhile, the special in-batch negative techniques \cite{DBLP:conf/emnlp/KarpukhinOMLWEC20} for DE allows it to assign more appropriate similarity scores for easy negatives, which is more suitable for the first stage training of the student model.
(2) Second, the performances of alone trained DE are limited. 
More specifically, when the student model distills to the later stage, it is difficult to quickly train a DE teacher model with better performance than the student model.

\paragraph{Distilling with CE Teachers.}
After learning with a 12-layer DE, we use two CEs as the teacher models.
Since CE can capture the fine-grained correlations between a pair of query and passage, it is more suitable for training with difficult data.
However, CE can not make an effective distinction between hard and easy negatives.
Such probability distributions of CE teachers are somehow confusing and inconducive for the student model.
Therefore, when deploying CE as the teacher model, we only use hard negatives in distillation.

In our preliminary study, we also find that it is critical to select a proper CE.
Generally speaking, the CE with more parameters can give more accurate predictions, but the more accurate predictions may not benefit distillation~\cite{DBLP:conf/aaai/MirzadehFLLMG20,DBLP:conf/iclr/ZhouGN20}. 
When the difficulties of training instances are not very conflating, the premature use of CE will lead to the performance loss of student model.
We think the main reason is that CE with high performances will output so confident results that the predicted probability distribution is often unfavorable to the student model.
Therefore, we perform 12-layer CE $\rightarrow$ 6-layer DE before 24-layer CE $\rightarrow$ 6-layer DE.

To overcome the catastrophic forgetting in learning with multiple teachers, we additionally follow \citet{DBLP:journals/pami/LiH18a} and \citet{DBLP:conf/emnlp/CaoCZW20}, using regularization to maintain the stability of training.
We first save a frozen copy of the student model as $S'$ and involve a \textbf{regularization loss} item $\Ls_r$ in distilling the student model $S$:
\begin{equation}
\footnotesize
\vd^{S'}_{de} = \frac{\exp(s_{de}^{S'}(\vq, \vp))}{\sum_{\vp' \in \{\vp^+, \sP^-\}} \exp(s_{de}^{S'}(\vq, \vp'))} \label{equ:studentcpoyscorepd}
\end{equation}
\begin{equation}
\footnotesize
\Ls_r(\vq,\vp^+,\sP^-)=\KL(\vd^{S'}_{de}/\tau, \vd^S_{de}/\tau) \label{equ:lwfsoftloss}
\end{equation}
where $s_{de}^{S'}$ is calculated by the copied student $S'$.

The overall loss function consists of three parts: the hard loss in \cref{equ:dtdhardloss}, the soft loss that adopts the relevant scores $s_{ce}^T$ calculated by the CE teacher model, and the regularization loss $\Ls_r$.
Finally, the loss function with a CE teacher is like:
\begin{equation}
\footnotesize
\vd^T_{ce} = \frac{\exp(s_{ce}^{T}(\vq, \vp))}{\sum_{\vp' \in \{\vp^+, \sP^-\}} \exp(s_{ce}^{T}(\vq, \vp'))} \label{equ:ceteacherscorepd}
\end{equation}
\begin{equation}
\footnotesize
\Ls_{ce \rightarrow de}(\vq,\vp^+,\sP^-)=\KL(\vd^T_{ce}/\tau, \vd^S_{de}/\tau) \label{equ:ctdsoftloss}
\end{equation}
\begin{equation}
\footnotesize
\Ls_\textrm{II} = \alpha_2 \Ls_h + \beta_2 \Ls_{ce \rightarrow de} + \gamma \Ls_r \label{equ:weightedloss2}
\end{equation}
where $\alpha_2$, $\beta_2$ and $\gamma$ control the relative importance.

\subsection{Data Progressive Distillation}
\label{sec:data_progressive}

After learning from progressively stronger teachers, the student model will be improved with knowledge distillation.
But there are still many confusing negative passages that lead to the disagreement between the teacher model and the student model, limiting the performance of the student model.
Therefore, we aim to fill the performance gap between the 24-layer CE teacher and the student model from the perspective of training data.

Intuitively, the data difficulty needs to be progressively raised to fit the capacity of the teacher, enhancing the student model in a clearer direction.
Therefore, sequentially after the last step of TPD, our solution is to adopt an iterative data selection procedure (let us say there are $N$ iterations), where each iteration consists of the following four steps:

(1) Retrieve the top-$k$ negatives with the current student.

(2) Collect the queries that the teacher model can predict the positive as top-$1$ but the student model can only predict as top-$k^\prime$. We construct a dataset $\sD^i$ for the $i$-th iteration as the collection of those queries, whose positive passages are the labeled ones and the negative passages are mined in step (1).

(3) Continually train the 24-layer CE teacher model with $\sD^i$.

(4) Use $\sD^i$ to distill with the teacher and the loss function in \cref{equ:weightedloss2}.

\section{Experiments}

\subsection{Experimental Setting}

We conduct experiments on several text retrieval datasets: 
MS MARCO Passage Ranking (\textsc{MS-Pas}) \cite{DBLP:conf/nips/NguyenRSGTMD16},
TREC 2019 Deep Learning Track (\textsc{TREC-Pas-19}, \textsc{TREC-Doc-19}) \cite{DBLP:journals/corr/abs-2003-07820},
MS MARCO Document Ranking (\textsc{MS-Doc}) \cite{DBLP:conf/nips/NguyenRSGTMD16},
Natural Questions (\textsc{NQ}) \cite{DBLP:journals/tacl/KwiatkowskiPRCP19},
and two industry datasets (\textsc{Bing-Rel} and \textsc{Bing-Ads}).
\textsc{Bing-Rel} contains multilingual positive and negative query-document pairs from Bing, where clicked documents are selected from search log data as positives and top-$k$ retrieval documents are mined against a fixed doc corpus as the hard negatives following ANCE \cite{DBLP:conf/iclr/XiongXLTLBAO21}.
We collect high quality click data from Bing ads for~\textsc{Bing-Ads} (here we only select ads with match type \textit{Phrase Match}~\footnote{\url{ https://help.ads.microsoft.com/apex/index/3/en-us/50822}}), the clicks resulting in user dwell time on landing pages for greater than 20 ms are considered as high quality clicks to exclude randomly accidental clicks. The corpus is formulated as ``query, ad keyword'', the ``ad keyword'' is the keyword corresponding to the clicked ad. we use the data from November 2021 to July 2022 as training set, the data from August 1, 2022 to August 10, 2022 and from August 10, 2022 to August 20, 2022 as validation set and test set, respectively.
The statistics are shown in \cref{sec:data}.

For \textsc{MS-Pas}, We follow the existing work \cite{DBLP:conf/emnlp/RenQLZSWWW21,DBLP:conf/sigir/ZhanM0G0M21}, reporting \textbf{MRR@10}, \textbf{Recall@50} and \textbf{Recall@1k} on the dev set.
For \textsc{TREC-Pas-19}, we select \textbf{nDCG@10} and \textbf{MAP@1k} as the evaluation metrics.
For \textsc{MS-Doc}, we report \textbf{MRR@10} and \textbf{Recall@100} on the dev set.
For \textsc{TREC-Doc-19}, we select \textbf{nDCG@10} and \textbf{Recall@100} as the evaluation metrics.
For \textsc{NQ}, we choose \textbf{Recall@5}, \textbf{Recall@20} and \textbf{Recall@100} as the evaluation metrics.
For \textsc{Bing-Rel} and \textsc{Bing-Ads}, we also treat \textbf{MRR@10}, \textbf{Recall@5}, \textbf{Recall@20} and \textbf{Recall@100} as the evaluation metrics. We conduct significant tests based on the paired t-test with $p\leq0.01$.


\subsection{Baselines}

We compare PROD with two groups of baselines.
The first group contains sparse retrieval methods and dense retrieval methods without knowledge distillation or multiple vectors, including 
BM25 \cite{DBLP:conf/sigir/Yang0L17},
DeepCT \cite{DBLP:conf/sigir/DaiC19},
docT5query \cite{DBLP:journals/corr/abs-1904-08375},
SPARTA \cite{DBLP:conf/naacl/ZhaoLL21}, 
GAR \cite{DBLP:conf/acl/MaoHLSG0C20},
DPR \cite{DBLP:conf/emnlp/KarpukhinOMLWEC20}, 
ANCE \cite{DBLP:conf/iclr/XiongXLTLBAO21}, 
RDR \cite{DBLP:journals/corr/abs-2010-10999},
Joint Top-$k$ \cite{DBLP:conf/acl/SachanPSKPHC20},
DPR-PAQ \cite{DBLP:conf/naacl/OguzLGLKPC0YGM22},
Ind Top-$k$ \cite{DBLP:conf/acl/SachanPSKPHC20},
STAR \cite{DBLP:conf/sigir/ZhanM0G0M21}, 
and ADORE \cite{DBLP:conf/sigir/ZhanM0G0M21}.

The second group is about dense retrieval methods producing a single vector for each document and query enhanced by knowledge distillation, including
Margin-MSE \cite{DBLP:journals/corr/abs-2010-02666},
TCT-ColBERT \cite{DBLP:journals/corr/abs-2010-11386},
TAS-B \cite{DBLP:conf/sigir/HofstatterLYLH21},
SPLADE v2 \cite{DBLP:journals/corr/abs-2109-10086},
RocketQA v1 \cite{DBLP:conf/naacl/QuDLLRZDWW21},
RocketQA v2 \cite{DBLP:conf/emnlp/RenQLZSWWW21},
PAIR \cite{DBLP:conf/acl/RenLQLZSWWW21},
and CL-DRD \cite{DBLP:conf/sigir/ZengZV22}.

\subsection{Implementation Details}

\begin{table*}[t]
\small
\centering
\begin{tabular}{l|c|ccc|cc}
\hline
\multirow{2}{*}{\textbf{Method}} &  \multirow{2}{*}{\textbf{\#Params}} & \multicolumn{3}{c|}{\textbf{\textsc{MS-Pas}}} & \multicolumn{2}{c}{\textbf{\textsc{TREC-Pas-19}}} \\
 & & \textbf{MRR@10} & \textbf{Recall@50} & \textbf{Recall@1k} & \textbf{nDCG@10} & \textbf{MAP@1k} \\ \hline
BM25 \cite{DBLP:conf/sigir/Yang0L17} &  - & 18.7 &  59.2 & 85.7 & 49.7 & 29.0 \\
DeepCT \cite{DBLP:conf/sigir/DaiC19} & - & 24.3 & 69.0 & 91 & 55.0 & 34.1 \\
docT5query \cite{DBLP:journals/corr/abs-1904-08375} & - & 27.2 & 75.6 & 94.7 & 64.2 & 40.3 \\
ANCE \cite{DBLP:conf/iclr/XiongXLTLBAO21} & 12-layer (110M) & 33.0 & - & 95.9 & 64.8 & 37.1 \\
ADORE \cite{DBLP:conf/sigir/ZhanM0G0M21} & 12-layer (110M) & 34.7 & - & - & 68.3 & 41.9 \\
\hline
TCT-ColBERT \cite{DBLP:journals/corr/abs-2010-11386} & 12-layer (110M) & 33.5 & - & 96.4 & 67.0 & 39.1 \\
RocketQA v1 \cite{DBLP:conf/naacl/QuDLLRZDWW21} & 12-layer (110M) & 37.0 & 85.5 & 97.9 & - & - \\
PAIR \cite{DBLP:conf/acl/RenLQLZSWWW21} & 12-layer (110M) & 37.9 & 86.4 & 98.2 & - & - \\
RocketQA v2 \cite{DBLP:conf/emnlp/RenQLZSWWW21} & 12-layer (110M) & 38.8 & 86.2 & 98.1 & - & - \\
Margin-MSE \cite{DBLP:journals/corr/abs-2010-02666} & 6-layer (66M) & 32.3 & - & 95.7 & 69.9 & 40.5 \\
TAS-B \cite{DBLP:conf/sigir/HofstatterLYLH21} & 6-layer (66M) & 34.4 & - & 97.6 & 71.7 & 44.7 \\
SPLADE v2 \cite{DBLP:journals/corr/abs-2109-10086} & 6-layer (66M) & 36.8 & - & 97.9 & 72.9 & - \\
CL-DRD \cite{DBLP:conf/sigir/ZengZV22} & 6-layer (66M) & 38.2 & - & - & 72.5 & 45.3 \\
\hline
PROD & 6-layer (66M) & \textbf{39.3}\bm{$^{\ast\dagger\ddagger\S}$} & \textbf{87.0}\bm{$^{\ast\dagger}$} & \textbf{98.4}\bm{$^{\ast\dagger\ddagger}$} & \textbf{73.3}\bm{$^{\ddagger\S}$} &  \textbf{48.4}\bm{$^{\S}$} \\
\hline
\end{tabular}
\caption{The main results on \textsc{MS-Pas} and \textsc{TREC-Pas-19}. ``\#Params'' represents the number of trainable parameters. We use the paired t-test with $p\leq0.01$. The superscripts refer to significant improvements compared to PAIR($^\ast$), RocketQA v2($^\dagger$), SPLADE v2($^\ddagger$), CL-DRD($^\S$).}
\label{tab:main_results1}
\end{table*}

\paragraph{Model Initialization.}
Following the settings of RocketQA v2 \cite{DBLP:conf/emnlp/RenQLZSWWW21}, both the 12-layer DE and the 12-layer CE use \textsc{ERNIE-2.0-Base} as the encoders for the public datasets and \textsc{Bing-Ads}.
Besides, the 24-layer CE uses \textsc{ERNIE-2.0-Large} as the encoder.
Please note that a well-pretrained checkpoint is not required for the student.
Therefore, we adopt the first six layers of \textsc{ERNIE-2.0-Base} as the 6-layer DE student for simplicity.
For \textsc{Bing-Rel}, we use \textsc{BERT-Multilingual-Base} as the encoders for 12-layer DE and 12-layer CE, simply taking the first six layers of \textsc{BERT-Multilingual-Base} as the student.

\begin{table}[t]
\small
\setlength\tabcolsep{3pt}
\begin{tabular}{l|ccc}
\hline
\textbf{Method} & \textbf{Recall@5} & \textbf{Recall@20} & \textbf{Recall@100} \\ \hline
BM25 \cite{DBLP:conf/sigir/Yang0L17} & - & 59.1 & 73.7 \\
GAR \cite{DBLP:conf/acl/MaoHLSG0C20} & 60.9 & 74.4 & 85.3 \\
DPR \cite{DBLP:conf/emnlp/KarpukhinOMLWEC20} & - & 78.4 & 85.4 \\
ANCE \cite{DBLP:conf/iclr/XiongXLTLBAO21} & - & 81.9 & 87.5 \\ 
RDR \cite{DBLP:journals/corr/abs-2010-10999} & - & 82.8 & 88.2 \\
Joint Top-$k$ \cite{DBLP:conf/acl/SachanPSKPHC20} & 72.1 & 81.8 & 87.8 \\
DPR-PAQ \cite{DBLP:conf/naacl/OguzLGLKPC0YGM22} & 74.5 & 83.7 & 88.6 \\
Ind Top-$k$ \cite{DBLP:conf/acl/SachanPSKPHC20} & 75.0 & 84.0 & 89.2 \\
\hline
RocketQA v1 \cite{DBLP:conf/naacl/QuDLLRZDWW21} & 74.0 & 82.7 & 88.5 \\
PAIR \cite{DBLP:conf/acl/RenLQLZSWWW21} & 74.9 & 83.5 & 89.1 \\
RocketQA v2 \cite{DBLP:conf/emnlp/RenQLZSWWW21} & 75.1 & 83.7 & 89.0 \\
\hline
PROD & \textbf{75.6}\bm{$^{\ast}$} & \textbf{84.7}\bm{$^{\ast\dagger}$} & \textbf{89.6}\bm{$^{\ast\dagger}$} \\
\hline
\end{tabular}
\caption{The main results on \textsc{NQ}. All the baselines are 12-layer, while the student in PROD is 6-layer. We use the paired t-test with $p\leq0.01$. The superscripts refer to significant improvements compared to Ind Top-k($^\ast$), RocketQA v2($^\dagger$).}
\label{tab:main_results2}
\end{table}


\paragraph{Distillation Warming Up.}
Before TPD, we need to warm up the first teacher.
We use random or BM25 negatives to train a 12-layer DE, retrieving the top-$k$ negatives, retraining a 12-layer DE with the mined negatives as the first teacher for \textsc{MS-Pas} and \textsc{MS-Doc}.
Especially, while retrieving the top-$k$ negatives for \textsc{NQ}, we additionally filter positive passages by matching the passages and answers.

\begin{table}[t]
\small
\centering
\setlength\tabcolsep{3pt}
\begin{tabular}{l|cc|cc}
\hline
\multirow{2}{*}{\textbf{Method}} & \multicolumn{2}{c|}{\textbf{\textsc{MS-Doc}}} & \multicolumn{2}{c}{\textbf{\textsc{TREC-Doc-19}}} \\ & \textbf{MRR@10} & \textbf{Recall@100} & \textbf{nDCG@10} & \textbf{Recall@100} \\ \hline
BM25 \cite{DBLP:conf/sigir/Yang0L17} & 27.9 & 80.7 & 51.9 & 39.5 \\
DeepCT\cite{DBLP:conf/sigir/DaiC19} & - & - & 54.9 & - \\
ANCE \cite{DBLP:conf/iclr/XiongXLTLBAO21}  & 37.7  & 89.4 & 61.0 & 27.3 \\
STAR \cite{DBLP:conf/sigir/ZhanM0G0M21}  & 39.0  & 91.3 & 60.5 & 31.3 \\
ADORE \cite{DBLP:conf/sigir/ZhanM0G0M21} & 40.5 & 91.9 & 62.8 & 31.7 \\
\hline
PROD  & \textbf{42.8}\bm{$^{\ast\dagger\S}$} & \textbf{93.3}\bm{$^{\ast\dagger\S}$} & \textbf{63.6}\bm{$^{\ast\S}$} & \textbf{32.6}\bm{$^{\ast\S}$} \\
\hline
\end{tabular}
\caption{The main results on \textsc{MS-Doc} and \textsc{TREC-Doc-19}. All baselines are 12-layer without distillation, while the student model in PROD is 6-layer.  We use the paired t-test with $p\leq0.01$. The superscripts refer to significant improvements compared to ANCE($^\S$), STAR($^\ast$), ADORE($^\dagger$).}
\label{tab:main_results3}
\end{table}

\paragraph{Hyper-parameter Setting.}
For mining the hard negatives, we choose $k=1000$ for the public datasets and $k=100$ for the industry datasets.
In DPD, we set $k^\prime=15$ and the iteration number $N=1$.
We use AdamW \cite{DBLP:conf/iclr/LoshchilovH19} as the optimizer.
Other detailed hyper-parameters for reproducing our experiments are shown in \cref{sec:hyps}.

\subsection{Main Results}

The results comparing PROD with the baselines on \textsc{MS-Pas}, \textsc{TREC-Pas-19}, \textsc{TREC-Doc-19}, \textsc{NQ} and \textsc{MS-Doc} are shown in \cref{tab:main_results1}, \cref{tab:main_results2}, \cref{tab:main_results3}, respectively.
It can be easily observed that PROD achieves state-of-the-art results in all five datasets, which can be concluded from two perspectives.

\paragraph{With or Without Distillation.}
As we can see from \cref{tab:main_results1} and \cref{tab:main_results2}, the baselines are divided into two groups, representing without and with distillation.
Comparing against ADORE (the best on \textsc{MS-Pas}, \textsc{TREC-Pas-19},  \textsc{MS-Doc} and \textsc{TREC-Doc-19} without distillation), Ind Top-$k$ (the best on \textsc{NQ} without distillation), RocketQA v2 (the best on \textsc{MS-Pas} and \textsc{NQ} with distillation) and CL-DRD (the best on \textsc{TREC-Pas-19} with distillation), PROD  can achieve better performances in all the groups.

\paragraph{6-layer or 12-layer.}
In \cref{tab:main_results1}, the performance of the best 12-layer method RocketQA v2 is better than the best 6-layer method CL-DRD on \textsc{MS-Pas}.
However, PROD further exceeds RocketQA v2 with a 6-layer architecture.
Besides, among all the 12-layer baselines on \textsc{NQ} and \textsc{MS-Doc}, PROD achieve the best results, showing the effectiveness of the progressive distillation method even with a 6-layer student.



\begin{table}[t]
\small
\centering
\begin{tabular}{l|ccc}
\hline
\textbf{Method} & \textbf{MRR@10} & \textbf{Recall@50} & \textbf{Recall@1k} \\ \hline
Pure Student & 31.66 & 79.99 & 96.19 \\ \hline
Random Batch & 37.94 & 85.82 & 98.09 \\
Merge Score & 37.64 & 84.93 & 97.68 \\
Merge Loss & 38.09 & 86.10 & 98.09 \\ \hline
TPD & 38.75 & 86.56 & 98.41 \\
TPD+DPD & \textbf{39.34} & \textbf{87.06} & \textbf{98.44} \\ \hline
\end{tabular}
\caption{The results of different methods with multiple teachers on \textsc{MS-Pas}.}
\label{tab:multiresult}
\end{table}

\subsection{Comparison with Multi-Teacher Methods}

To further prove the effectiveness of our method, we compare PROD with three methods that also incorporated with multiple teachers.
Random Batch \cite{DBLP:conf/interspeech/FukudaSKTCR17} randomly selects a teacher in each batch of training.
Merge Score \cite{DBLP:conf/interspeech/FukudaSKTCR17} averages the soft labels of multiple teachers in training.
And Merge Loss \cite{DBLP:conf/interspeech/FukudaSKTCR17} adds up all the distillation loss of each teacher, before calculating gradients.
Please note that all the methods use the same set of teachers as PROD, \textit{i.e.}, 12-layer DE, 12-layer CE and 24-layer CE.
We also report the results of training a 6-layer DE without distillation, which is denoted as ``Pure Student''.

As we can see from the results of \textsc{MS-Pas} in \cref{tab:multiresult}, all the multi-teacher methods can lead to performance gain comparing with Pure Student.
Besides, PROD can achieve the best performances among these multi-teacher methods even just uses TPD, which also reflects the effectiveness of our method.

\begin{table*}[t]
\small
\centering
\setlength\tabcolsep{3pt}
\resizebox{\textwidth}{!}{
\begin{tabular}{l|l|c|ccccccc}
\hline
\textbf{Dataset} & \textbf{Teacher Variant} & \textbf{Student} & \textbf{MRR@10} & \textbf{Recall@1} & \textbf{Recall@5} & \textbf{Recall@20} & \textbf{Recall@50} & \textbf{Recall@100} & \textbf{Recall@1k} \\
\hline
\multirow{17}{*}{\textbf{\textsc{MS-Pas}}} & - & 6DE & 31.66 & 20.14 & 47.45 & 68.95 & 79.99 & 85.59 & 96.19 \\
& \textbf{12DE}$^*$ & 6DE & 35.69 & 23.65 & 52.54 & 73.94 & 84.09 & 89.31 & 97.59 \\
& 12CE & 6DE & 37.44 & 25.11 & 54.11 & 75.03 & 84.44 & 88.81 & 97.15 \\
& 24CE & 6DE & 36.18 & 23.80 & 53.08 & 74.57 & 84.15 & 89.13 & 97.46 \\
& \textbf{12DE->12CE}$^*$ & 6DE & 38.09 & 25.34 & 55.54 & 76.75 & 86.12 & 90.83 & 97.72 \\
& 12CE->12DE & 6DE & 34.90 & 22.98 & 51.07 & 73.01 & 82.95 & 88.44 & 97.22 \\
& 12DE->24CE & 6DE & 38.08 & 25.46 & 55.46 & 76.76 & 85.72 & 90.53 & 97.82 \\
& \textbf{12DE->12CE->24CE}$^*$ & 6DE & 38.75 & 25.89 & 55.86 & 77.68 & 86.56 & 91.33 & 98.41 \\ 
& 12CE->12CE->24CE & 6DE & 38.29	& 25.87 & 55.14 & 75.64 & 84.58 & 88.62 & 96.40 \\
& 12DE->12CE->12CE & 6DE & 37.88 & 24.94 & 55.53 & 76.62 & 86.12 & 90.76 & 98.01 \\
& 12DE->24CE->24CE & 6DE & 37.63 & 24.87 & 54.99 & 76.39 & 85.54 & 90.26 & 98.07 \\
& \textbf{12DE->12CE->24CE->DTD}$^*$ \textbf{(PROD)} & 6DE & 39.34 & 26.66 & 56.35 & 78.09 & 87.06 & 91.52 & 98.44 \\
& 12DE->12CE->24CE->24CE & 6DE & 37.98 & 25.06 & 55.32 & 77.06 & 85.93 & 90.50 & 97.54 \\
\cline{2-10}
& - & 2DE & 27.31 & 17.18 & 40.83 & 61.65 & 72.79 & 79.79 & 93.57 \\
& 12DE & 2DE & 30.78 & 19.27 & 46.48 & 67.87 & 78.88 & 85.32 & 96.29 \\
& 12DE->12CE & 2DE & 34.10 & 22.41 & 50.59 & 71.15 & 80.62 & 86.32 & 95.56 \\
& 12DE->24CE & 2DE & 32.92 &	21.68 & 48.47 & 69.84 & 80.24 & 85.95 & 95.49 \\
\hline \hline
\multirow{8}{*}{\textbf{\textsc{NQ}}} & - & 6DE & - & 45.96 & 67.76 & 78.88 & - & 86.24 & - \\
& \textbf{12DE}$^*$ & 6DE & - & 54.82 & 73.73 & 82.70 & - & 88.29 & - \\
& 12CE & 6DE & - & 52.16 & 71.54 & 81.78 & - & 88.24 & - \\
& 24CE & 6DE & - & 51.19 & 70.57 & 80.62 & - & 87.15 & - \\
& \textbf{12DE->12CE}$^*$ & 6DE & - & 57.12 & 75.12 & 84.30 & - & 88.82 & - \\
& 12DE->24CE & 6DE & - & 56.87 & 74.75 & 84.22 & - & 88.81 & - \\
& \textbf{12DE->12CE->24CE}$^*$ & 6DE & - & 57.20 & 75.54 & 84.66 & - & 89.48 & - \\
& \textbf{12DE->12CE->24CE->DTD}$^*$ \textbf{(PROD)} & 6DE & - & 57.63 & 75.61 & 84.72 & - & 89.56 & - \\ \hline
\end{tabular}
}
\caption{The results of ablation study on \textsc{MS-Pas} and \textsc{NQ}. ``$^*$'' means the actual steps in PROD. For simplicity, we use ``$n$DE'' and ``$n$CE'' to denote the $n$-layer DE and $n$-layer CE, respectively. We use ``A->B'' to denote continual distillation using A and B as the teachers in turn with refreshed top-$k$ negatives before learning from each teacher.}
\label{tab:ablation}
\end{table*}

\begin{table*}[t]
\small
\centering
\begin{tabular}{l|c|ccccccc}
\hline
\textbf{Dataset} & \textbf{Teacher} & \textbf{MRR@10} & \textbf{Recall@1} & \textbf{Recall@5} & \textbf{Recall@20} & \textbf{Recall@50} & \textbf{Recall@100} & \textbf{Recall@1k} \\ \hline
\multirow{3}{*}{\textbf{\textsc{MS-Pas}}} & 12DE & 35.77 & 23.50 & 52.72 & 74.18 & 84.10 & 89.14 & 97.61 \\
& 12CE & 40.81 & 27.88 & 58.58 & 79.04 & 86.52 & 90.81 & 97.61 \\
& 24CE & 41.96 & 28.84 & 59.81 & 80.00 & 87.89 & 92.23 & 97.61 \\
\hline \hline
\multirow{3}{*}{\textbf{\textsc{NQ}}} & 12DE & - & 52.94 & 73.15 & 82.70 & - & 88.68 & - \\
& 12CE & - & 60.38 & 79.46 & 86.02 & - & 88.68 & - \\
& 24CE & - & 64.78 & 80.57 & 86.24 & - & 88.68 & - \\ \hline
\end{tabular}
\caption{The performance of the teachers in the first step of TPD on \textsc{MS-Pas} and \textsc{NQ}. For simplicity, we use ``$n$DE'' and ``$n$CE'' to denote the $n$-layer DE and $n$-layer CE, respectively. The performances of CEs are the reranking results based on the retrieval output of the 12-layer DE.}
\label{tab:ablation_step1_additional}
\end{table*}

\subsection{Ablation Study}

To investigate the effectiveness of TPD and DPD, we conduct a careful ablation study on the two parts.
Please note that we use ``12DE->12CE->24CE->DTD'' to denote our method PROD, showing the specific steps inside. 

\paragraph{Effect of TPD}
The performances of each distillation step on \textsc{MS-Pas} and \textsc{NQ} are shown in \cref{tab:ablation}.
Compared with the original 6-layer DE without distillation, the performance of ``12DE->12CE->24CE->DTD'' on \textsc{MS-Pas} has been improved by about 7.7\% in MRR@10, 7.1\% in Recall@50 and 2.2\% in Recall@1k, proving the overall effectiveness of PROD method.
In addition, by comparing with the actual teacher variants used in PROD, we can easily find that each step of PROD has achieved stable improvement.
We focus on the selection of the teacher model in each step and conduct further experiments to prove the correctness and necessity of the teacher's order in each step of PROD.

$\bullet$ \textbf{Influence of Teacher \#1.}
How to choose Teacher \#1 is a problem worth exploring.
Therefore, in the first step of teacher progressive, we use three different teachers, ``12DE'', ``12CE'' and ``24CE'', to distill 6-layer DE.
Although the experimental results show that ``12DE'' on \textsc{MS-Pas} is worse than ``12CE'' and ``24CE'' for MRR@10 and Recall@100, there are several important reasons why we still choose 12-layer DE as Teacher \#1.

(1) Comparing ``12DE'' with ``12CE'' and ``24CE'', Recall@50 on \textsc{MS-Pas} are higher than CEs.
We believe that this is because DE uses in-batch negatives, which increases the overall understanding of retrieval task, and is helpful for the student model to further learn more difficult knowledge.

(2) Compared with more later distillation steps, it is surprising that we can not get better results by adjusting the order of DE distillation backward or completely abandoning DE.
It shows that DE distillation mainly increases the model's cognition of easy negatives.
With the increase of data difficulty and student model performance in the training, it is too late to use DE distillation.

(3) In \cref{tab:ablation_step1_additional}, we can see the performance of different teacher models on \textsc{MS-Pas} and \textsc{NQ}, where CE performances are measured by reranking the results of DE.
Although the performance of 12-layer DE is the worst, the distillation efficiency is the highest, that is, the student model is the closest to the teacher's performance after distillation.
This phenomenon is more obvious on \textsc{NQ} shown in \cref{tab:ablation}, where the student model can perform best by using 12-layer DE as Teacher \#1, while the performance of the 12-layer DE is the worst compared with CE.
These results further support that the selection of distillation teachers at different steps should not take the performance of teacher models as the only standard.

$\bullet$ \textbf{Influence of Teacher \#2.}
In the second and third steps of teacher progressive, we adopt 12-layer CE and 24-layer CE as teacher models to distill student model.
As shown in \cref{tab:ablation}, using 24-layer CE as Teacher \#2 is almost the same or even worse than 12-layer CE.
We believe that this is caused by the excessive gap between teacher and student.
To verify this opinion, we increase the gap between teacher and student, repeated the experiment of the second step on a 2-layer DE student model.
The results are shown in \cref{tab:ablation}, which makes the comparison of taking 24-layer CE or 12-layer CE as the teacher in the second step more obvious.

Moreover, in order to explore whether 12-layer CE distillation can also perform well in the third step, we replaced the 24-layer CE in the third step with 12-layer CE, which is denoted as ``12DE->12CE->12CE''.
By comparing with ``12DE->12CE->24CE'', we can see that 24-layer CE, a more powerful teacher, must be used in the third step to achieve the best results, and this process must be transited from the 12-layer CE distillation.

Specifically, by comparing ``12DE->12CE->24CE'' with ``12DE->24CE->24CE'', we can see that replacing the 12-layer CE with the 24-layer CE in the second step for distillation cannot achieve satisfactory performance.
Therefore, it is necessary and reasonable for us to adopt two different architectures of CE, which are a 12-layer CE teacher in the second distillation step and a 24-layer CE teacher in the third distillation step in TPD.

\paragraph{Effect of DPD}
We also care about how much DPD contributes to our framework PROD.
By comparing ``12DE->12CE->24CE->DTD'' with ``12DE->12CE->24CE'' on \textsc{MS-Pas} and \textsc{NQ}, we can see that appending DTD after TPD can lead to consistent improvements on all the evaluation metrics used in both \textsc{MS-Pas} and \textsc{NQ}, showing the necessity of continual distillation using the confusing negative passages.
Further more, to verify whether the performance gain comes from the confusing data, we design another experiment by continual distillation on all the queries with the refreshed hard negatives using 24-layer CE teacher, which is denoted as ``12DE->12CE->24CE->24CE''.
By comparing ``12DE->12CE->24CE->DTD'' with ``12DE->12CE->24CE->24CE'', we can observe performance improvements on \textsc{MS-Pas}.
We think the reason may be that the noises in the entire training set obstructs the student's learning, which also means that mining a small amount of confusing data in DPD is necessary and improves the efficiency of distillation.

\begin{table}[t]
\small
\centering
\setlength\tabcolsep{2pt}
\begin{tabular}{c|l|c|cccccccc}
\hline
\textbf{Dataset} & \textbf{Teacher} & \textbf{Student} & \textbf{MRR@10} & \textbf{R@5} & \textbf{R@20} & \textbf{R@100} \\
\hline
\multirow{6}{*}{\rotatebox{90}{\textbf{\textsc{Bing-Rel}}}}
& - & 6DE & 39.51 & 54.27 & 70.78 & 83.02 \\
& 12DE & 6DE & 42.46 & 56.82 & 73.81 & 85.92 \\
& 12DE->12CE & 6DE & 43.81 & 58.80 & 74.18 & 85.72 \\
& PROD & 6DE & 44.37 & 59.71 & 74.72 & 85.79 \\
\cline{2-7}
& 12DE & - & 42.51 & 57.71 & 73.89 & 85.81 \\
& 12CE & - & 49.92 & 66.21 & 80.45 & 85.81 \\
\hline \hline
\multirow{6}{*}{\rotatebox{90}{\textbf{\textsc{Bing-Ads}}}}
& - & 6DE & 22.30 & 33.54 & 57.39  & 82.44 \\
& 12DE & 6DE & 22.98  & 34.32 & 57.90 & 82.80  \\
& 12DE->12CE & 6DE & 24.07 & 35.54 & 59.24 & 83.71 \\
& PROD & 6DE & 24.82 & 36.93 & 60.47 & 84.30 \\
\cline{2-7}
& 12DE & - & 23.46 & 35.25 & 59.59 & 84.32 \\
& 12CE & - & 25.37 & 37.18 & 60.02 & 84.32 \\
\hline
\end{tabular}
\caption{The results on two industry datasets \textsc{Bing-Rel} and \textsc{Bing-Ads}. R@$n$ is short for Recall@$n$.
We use ``$n$DE'' and ``$n$CE'' to denote the $n$-layer DE and $n$-layer CE, respectively. We use ``A->B'' to denote continual distillation using A and B as the teachers in turn with refreshed top-$k$ negatives before learning from each teacher. The performances of CEs are based on the output of the 12-layer DE.}
\label{tab:industry}
\end{table}
\subsection{Results on Industry Datasets}

We conduct experiments on the two industry datasets by comparing each steps in PROD.
Please note that the scales of the industry datasets are larger than those of the public datasets.
Therefore, considering the cost of computational resources to reach the model's convergence, we only use 12-layer DE and 12-layer CE in PROD.
For \textsc{Bing-Rel}, we simply evaluate the last checkpoint after training and report the results on the dev set.
The results on \textsc{Bing-Rel} and \textsc{Bing-Ads} are illustrated in \cref{tab:industry}.

By comparing the results of each step in PROD in the two industry datasets, it is obvious that the performances increase consistently when more teachers or DPD are applied.
It indicates that PROD can lead to significant performance gain in practical industrial scenarios.
Additionally, after distilling using 12-layer CE, the performances of the 6-layer DE student can outperform the 12-layer DE teacher.
Besides, we can see that the performances of the 12-layer CE teacher are much better than the student even after distilling with that teacher.
But if applying DPD afterward, the performance of the student is much closer to the 12-layer CE teacher, which supports the fact that PROD has the potential to better alleviate the gap between the teacher and the student.

\section{Intrinsic Evaluation}

We explore the intrinsic properties of PROD from several aspects.

\begin{table}[t]
\small
\centering
\begin{tabular}{l|c|cc}
\hline
\textbf{Method} & \textbf{MRR@10} & \textbf{Recall@50} \\ \hline
TPD & 38.75 & 86.56 \\ 
\hline
ST-(1,2] \& TT-(0,1] & 39.13 & 86.89 \\
ST-(1,5] \& TT-(0,1] & 39.26 & 87.02 \\
ST-(1,15] \& TT-(0,1] & \textbf{39.34} & 87.06 \\
ST-(5,20] \& TT-(0,5] & 39.29 & 86.82 \\ 
ST-(0,15] < TT-(0,15] & 39.30 & 86.98 \\
ST-(0,31] < TT-(0,31] & 39.13 & \textbf{87.12} \\ \hline
\end{tabular}
\caption{The impact of different strategies of selecting the confusing negative passages on \textsc{MS-Pas}. ``ST'' and ``TT'' means student's and teacher's top ranking passages. respectively. ``A \& B'' denotes the intersection of A and B. ``A < B'' represents the passages that have higher relevance scores in B than A.}
\label{tab:filtertype}
\end{table}

\subsection{Confusing Negative Selection in DPD}

We first explore the impact of different confusing negative passages selection strategies in DPD.
All experiments are based on the same student model after TPD, the results are shown in \cref{tab:filtertype}.
The experiment results show that different selection strategies can improve the student model performance to a certain extent.
Besides, all the strategies outperform TPD.
Among them, ``ST-(1,15] \& TT-(0,1]'' performs best in MRR@10, which is the strategy we finally use to select the confusing data.

\begin{table}[t]
\small
\setlength\tabcolsep{3pt}
\centering
\begin{tabular}{c|cccc}
\hline
\textbf{Setting} & \textbf{MRR@10} & \textbf{Recall@5} & \textbf{Recall@20} & \textbf{Recall@50} \\ \hline
$N = 0$ & 38.75 & 55.86 & 77.68 & 86.56 \\
$N = 1$ & 39.34 & 56.35 & 78.09 & 87.06 \\
$N = 2$ & \textbf{39.43} & \textbf{56.39} & 77.95 & 86.89 \\
$N = 3$ & 39.28 & 56.36 & \textbf{78.14} & \textbf{87.09} \\
$N = 4$ & 39.17 & 56.38 & 78.11 & 86.93 \\
$N = 5$ & 39.08 & 56.16 & 78.11 & 86.95 \\
\hline
\end{tabular}
\caption{The results of different iteration number $N$ of DPD on \textsc{MS-Pas}.}
\label{tab:iterresult}
\end{table}

\subsection{Iteration Number of DPD}

In this section, we discuss the impact of different iteration number $N$ of DPD by gradually increasing it. 
As results shown in \cref{tab:iterresult}, when the iteration is at early stage, the improvement of DPD is obvious.
Specifically, MRR@10 reaches the best 37.44 when $N = 2$ and Recall@50 reaches the best 87.09 when $N = 3$.
However, when the iteration number is bigger than 3, the performance of the student model decreases.
We think the reason may be the long training steps of a small amount of data, which leads to the inevitable knowledge forgetting and overfitting.
Therefore, considering the training steps and the average performance, we choose $N = 1$.

\begin{table}[t]
\small
\centering
\begin{tabular}{l|l|ccc}
\hline
\textbf{Dataset} & \textbf{Method} & \textbf{MRR@10} & \textbf{Recall@50} & \textbf{Recall@1k} \\ \hline
\multirow{2}{*}{\textbf{\textsc{MS-Pas}}} & PROD & \textbf{39.34} & \textbf{87.06} & \textbf{98.44} \\
& PROD w/o $\Ls_r$ & 39.20 & 86.93 & 98.31 \\ \hline \hline
\textbf{Dataset} & \textbf{Method} & \textbf{Recall@5} & \textbf{Recall@20} & \textbf{Recall@100} \\ \hline
\multirow{2}{*}{\textbf{\textsc{NQ}}} & PROD & \textbf{75.61} & \textbf{84.72} & \textbf{89.56} \\
& PROD w/o $\Ls_r$ & 75.30 & 84.52 & 89.45 \\
\hline
\end{tabular}
\caption{The impact of the regularization loss item $\Ls_r$ in PROD on \textsc{MS-Pas} and \textsc{NQ}. }
\label{tab:LwFresult}
\end{table}

\subsection{Regularization Loss Item $\Ls_r$}

Last but not least, we explore the influence of the regularization loss item $\Ls_r$ in distillation.
Since the essence of the regularization loss item is to use the model in the previous step as a teacher and distill the current students, it is particularly effective in the situation where training data is scarce or the training is unstable and easy to overfit.
In order to observe the effect of the regularization loss item more intuitively, we show the results on \textsc{MS-Pas} and \textsc{NQ} in \cref{tab:LwFresult}.
In the experiment, we compare between the distillation results with and without the regularization loss item. 
The results show that the regularization loss item not only stabilizes the model performances but also improves distillation effect, effectively alleviating the trend of overfitting in distillation training.

\section{Conclusion}
In this paper, we propose a novel distillation method PROD for dense retrieval.
Concretely, we design \textit{teacher progressive distillation} and \textit{data progressive distillation} to gradually improve the performance of the student model.
Extensive experiments on five widely-used benchmarks show that PROD can effectively improve the performance of the student model, achieving new state-of-art within the the distillation methods for dense retrieval, even surpassing some existing 12-layer models.

\bibliographystyle{ACM-Reference-Format}
\bibliography{ref}

\newpage
\appendix
\section{Data Statistics}
\label{sec:data}

\begin{table}[t]
\centering
\small
\begin{tabular}{l|cccc}
\hline
\textbf{Dataset} & \textbf{Train} & \textbf{Dev} & \textbf{Test} & \textbf{\#Doc} \\
\hline
\textsc{MS-Pas} & 502,939 & 6,980 & - & 8,841,823 \\
\textsc{TREC-Pas-19} & - & - & 200 & 8,841,823 \\
\textsc{MS-Doc} & 367,013 & 5,193 & - & 3,213,835 \\
\textsc{TREC-Doc-19} & - & - & 200 & 3,213,835 \\
\textsc{NQ} & 58,880 & 8,757 & 3,610 & 21,015,324 \\
\textsc{Bing-Rel} & 1,593,219 & 8,013 & - & 5,335,927 \\
\textsc{Bing-Ads} & 8,306,968 & 53,219 & 52,590 & 2,866,527  \\
\hline
\end{tabular}
\caption{Statistics of the text retrieval datasets.} 
\label{tab:dataset}
\end{table}

The statistics of used datasets are shown in \cref{tab:dataset}.

\section{Hyper-parameters}
\label{sec:hyps}

\begin{table}[t]
\centering
\small
\setlength\tabcolsep{1pt}
\resizebox{\columnwidth}{!}{
\begin{tabular}{ll|l|ccccc}
\hline
 & & \textbf{Parameter} & \textbf{\textsc{NQ}} & \textbf{\textsc{MS-Pas}} & \textbf{\textsc{MS-Doc}} & \textbf{\textsc{Bing-Rel}} & \textbf{\textsc{Bing-Ads}} \\ \hline
\multicolumn{2}{l|}{\multirow{6}{*}{\textbf{Global}}} & Max query len & 32 & 32 & 32 & 64 & 32 \\
 & & Max passage len & 128 & 144 & 480 & 512 & 32 \\
 & & Temperature & 4.0 & 4.0 & 4.0 & 4.0 & 4.0 \\
 & & Hard loss weight & 0.1 & 0.1 & 0.1 & 0.1 & 0.1\\
 & & Soft loss weight & 0.9 & 0.9 & 0.9 & 0.9 & 0.1\\
 & & Warmup & 0.1 & 0.1 & 0.1 & 0.1 & 0.1 \\ \hline
\multicolumn{1}{l|}{\multirow{8}{*}{\textbf{TPD}}} & \multirow{4}{*}{\textbf{DE}} & Learning rate & 5e-5 & 5e-5 & 5e-5 & 2e-5 & 5e-5  \\
\multicolumn{1}{l|}{} & & Batch size & 128 & 128 & 128 & 256 & 512 \\
\multicolumn{1}{l|}{} & & Distillation step & 80000 & 40000 & 40000 & 16000 & 40000  \\
\multicolumn{1}{l|}{} & & Negative num & 1 & 1 & 1 & 2 & 1 \\
\cline{2-8} 
\multicolumn{1}{l|}{} & \multirow{4}{*}{\textbf{CE}} & Learning rate & 5e-5 & 5e-5 & 5e-5 & 2e-5 & 5e-5 \\
\multicolumn{1}{l|}{} & & Batch size & 64 & 64 & 64 & 64 & 256 \\
\multicolumn{1}{l|}{} & & Distillation step & 20000 & 40000 & 40000 & 16000 & 16000  \\
\multicolumn{1}{l|}{} & & Negative num & 15 & 15 & 15 & 15 & 15  \\
\hline
\multicolumn{2}{l|}{\multirow{4}{*}{\textbf{DPD}}} & Learning rate & 1e-5 & 1e-5 & 1e-5  & 1e-5 & 1e-5 \\
 & & Batch size & 64 & 64 & 64 & 64 & 256 \\
 & & Distillation step & 200 & 2000 & 2000  & 2000 & 2000  \\
 & & Negative num & 15 & 15 & 15 & 15 & 15 \\
\hline
\end{tabular}
}
\caption{Hyper-parameters for PROD.}
\label{tab:hyps}
\end{table}

The detailed hyper-parameters are shown in \cref{tab:hyps}.

\end{document}